\documentclass[lettersize,journal]{IEEEtran}
\IEEEoverridecommandlockouts
\usepackage{cite}
\usepackage{amsmath,amssymb,amsfonts}
\usepackage{algorithm,algorithmic}
\usepackage{graphicx}
\usepackage{url}
\graphicspath{{images/},{.}}
\DeclareGraphicsExtensions{.pdf,.jpeg,.png,.jpg}
\usepackage{textcomp}
\usepackage{verbatim}
\usepackage{xcolor}
\usepackage{multirow,multicol}

\def\BibTeX{{\rm B\kern-.05em{\sc i\kern-.025em b}\kern-.08em
    T\kern-.1667em\lower.7ex\hbox{E}\kern-.125emX}}

\usepackage{float}
\newfloat{footnote}{hb}

\begin{document}
\title{Fully-Binarized, Parallel, RRAM-based Computing Primitive for In-Memory Similarity Search\vspace{-0.5em}}
\author{\IEEEauthorblockN{Sandeep Kaur Kingra$^{1*}$, Vivek Parmar$^{1*}$, Deepak Verma$^{1}$, Alessandro Bricalli$^2$, Giuseppe Piccolboni$^2$,\\ Gabriel Molas$^2$, Amir Regev$^2$, and Manan Suri$^{1\dagger}$\vspace{-2em}}
\thanks{This work was supported in part by SERB-CRG/2018/001901 and CYRAN AI Solutions. Authors would also like to thank J.F. Nodin and G. Pares (CEA-LETI, Grenoble) for their contribution to ReRAM device fabrication at CEA-LETI R\&D facility in Grenoble.
$^1$S. K. Kingra, V. Parmar, D. Verma and M. Suri are with the Indian Institute of Technology Delhi, New Delhi. ($^\dagger$Corresponding Author: Manan Suri, e-mail: manansuri@ee.iitd.ac.in).
$^2$A. Bricalli, G. Piccolboni, G. Molas and A. Regev are with Weebit Nano, Hod Hasharon, Israel. At the time of the start of this work, the affiliation of G. Molas was CEA-LETI, Grenoble. $^*$S. K. Kingra and V. Parmar contributed equally for the manuscript.}}
\maketitle

\begin{abstract}
In this work, we propose a fully-binarized XOR-based IMSS (In-Memory Similarity Search) using RRAM (Resistive Random Access Memory) arrays. XOR (Exclusive OR) operation is realized using 2T-2R bitcells arranged along the column in an array. This enables simultaneous match operation across multiple stored data vectors by performing analog column-wise XOR operation and  summation to compute HD (Hamming Distance). The proposed scheme is experimentally validated on fabricated RRAM arrays. Full-system validation is performed through SPICE simulations using open source Skywater 130 nm CMOS PDK demonstrating energy of 17 fJ per XOR operation using the proposed bitcell with a full-system power dissipation of 145 $\mu$W. Using projected estimations at advanced nodes (28 nm) energy savings of $\approx$1.5$\times$ compared to the state-of-the-art can be observed for a fixed workload. Application-level validation is performed on HSI (Hyper-Spectral Image) pixel classification task using the Salinas dataset demonstrating an accuracy of 91\%. 
\end{abstract}


\section{Introduction}
Associative memories (or CAM - Content Addressable Memory) are an important component of intelligent systems that can perform fast search operations \cite{PanFSP21}. CAMs accept a query and perform search over multiple data points stored in memory to find one or more matches based on a distance metric and return locations of matches. This information can be potentially used for applications such as nearest neighbour searches for classification or unsupervised labeling\cite{ImaniMWPR19,Wu_2018,Imani_2020,Li_2021}. One of the basic distance metrics that can be used for such applications is HD (Hamming Distance)\cite{ImaniMWPR19}. For any pair of strings or words of equal length, HD is defined as the total number of positions where the symbols/characters of the pair differ from each other. Conventional CAMs are designed using standard memory technologies such as SRAM (16T \cite{MohanS09}, 9T NOR-based and 10T NAND-based bitcells \cite{PagiamtzisS06}) or DRAM \cite{Noda_2005,Lines}. However, such volatile memory-based circuits have performance limitations that can be potentially addressed by using emerging NVM (Non-Volatile Memory) devices\cite{7159147,Sun_2018,Shreya_2020}. Use of NVM devices provides additional design flexibility by reducing circuit complexity and providing opportunity to exploit low-area analog IMC (In-Memory Computing)\cite{7159147,Ielmini_2018}. Associative memory architectures that exploit NVM based IMC have been recently demonstrated using RRAM (Resistive RAM) devices based on XNOR\cite{Matsunaga_2014,Li_2014,Ni_2019,Yang_2019,Li_2021,PanFSP21} and XOR\cite{Wu_2018,HalawaniLMAA18} logic using HD as the distance metric. In this work, an end-to-end scheme is proposed to realize IMSS (In-Memory Similarity Search) in hardware by using RRAM devices and binarizing data and queries through a custom pre-processing pipeline. {XOR gate functionality is realized using 2T-2R RRAM circuits where one input is encoded in form of RRAM device conductance states and the other input is applied as voltage signals. Reasons for adopting the 2T-2R bitcell for the study are: (i) Improved tolerance for D2D (Device-to-Device)/C2C (Cycle-to-Cycle) variability compared to 1T-1R resulting in reliable operations even at relaxed programming conditions potentially increasing endurance\cite{Bocquet_2018}; (ii) Improved resilience to impact of resistance drift or read-disturbs for the programmed states and (iii) High signal margins easing the sensing requirements\cite{Yang_2019}. 
\begin{footnote}[!b]
© 2022 IEEE. Personal use of this material is permitted.  Permission from IEEE must be obtained for all other uses, in any current or future media, including reprinting/republishing this material for advertising or promotional purposes, creating new collective works, for resale or redistribution to servers or lists, or reuse of any copyrighted component of this work in other works.
\end{footnote}
\begin{figure}[tb]
  \centering
  \includegraphics[width=0.85\linewidth]{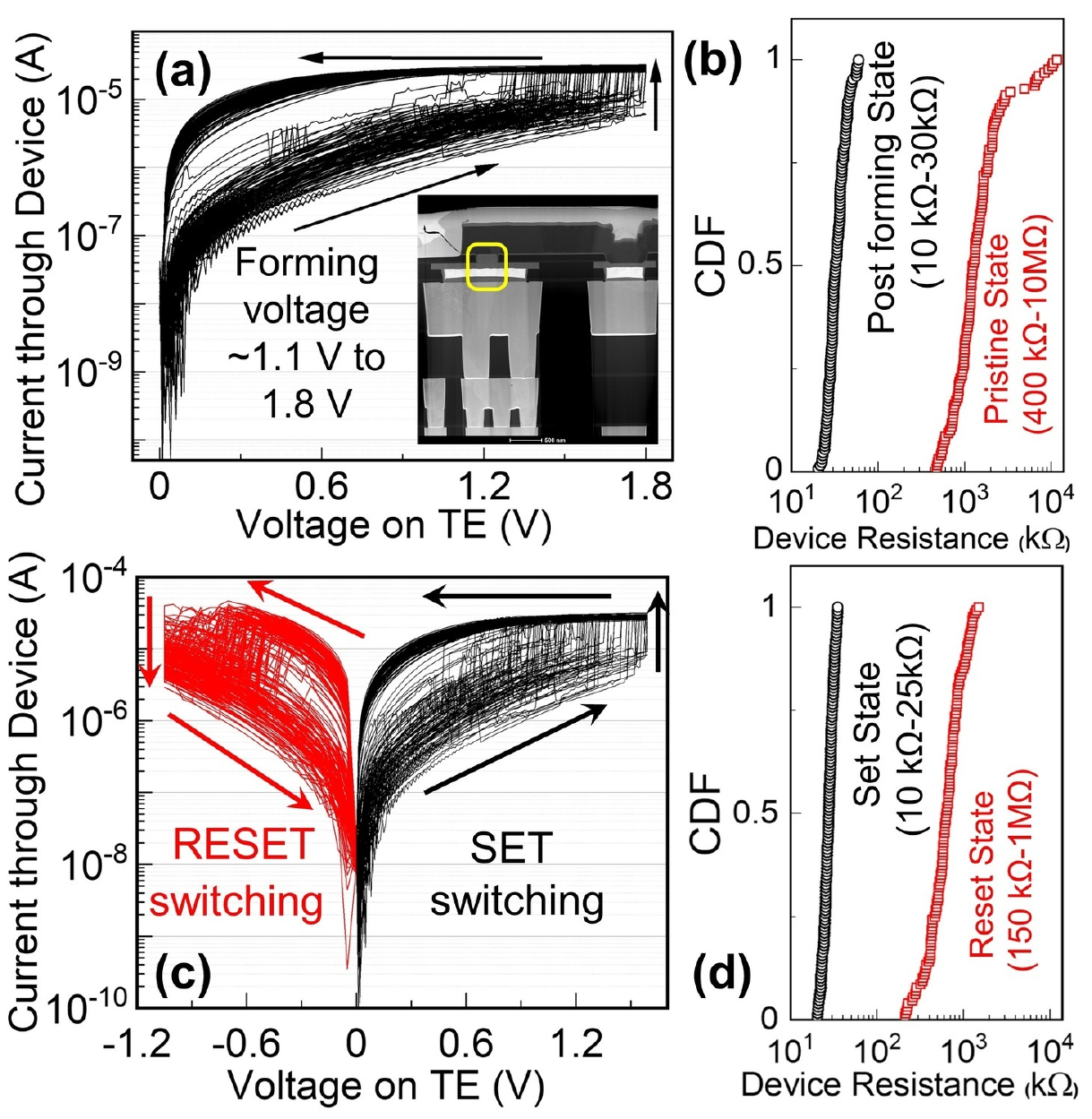}
  \vspace{-1em}
  \caption{(a) IV characteristics showing electro-forming [Inset: SEM cross-section of the SiO$_x$ RRAM cell integrated on top of the 130 nm CMOS], (b) Statistical resistance state distribution for pristine and post-forming resistance state (64 devices), (c) IV characteristics showing SET and RESET switching operation highlighting D2D variability, (d) Statistical resistance state distribution for LRS and HRS (64 devices). \vspace{-0.5em}}
  \label{Fig1}
\end{figure}
In an array structure when such circuits are arranged in column-orientation, QI (Query Input) can be simultaneously applied to multiple columns of SD (Stored Data). Key contributions of the work are: (i) Experimental validation of 2T-2R XOR bitcell operation on fabricated 8$\times$8 1T-1R RRAM arrays, (ii) Scheme for performing fully-binarized XOR-based IMSS through analog computation of HD, (iii) Validation of IMSS peripheral circuits through detailed SPICE simulations using 130 nm Skywater PDK, (iv) Validation of proposed methodologies on HSI (Hyper-Spectral Image) classification application with Salinas dataset and (v) Analysis depicting impact of array size on sensing margin and variability on classification accuracy. The manuscript is organized as follows: Section \ref{sec2} summarizes the RRAM device fabrication flow. Section \ref{sec3} presents the experimental validation of proposed RRAM IMSS scheme. Section \ref{sec4} presents the SPICE simulation results for IMSS architecture considering the periphery blocks. Section \ref{sec5} presents the application level validation of the proposed IMSS scheme on HSI pixel classification task using the Salinas dataset and finally Section \ref{sec6} provides concluding remarks.

\section{Fabricated RRAM Array}
\label{sec2}
The fabricated test chip used in our study is of size 16kb consisting of 256 8$\times$8 1T-1R RRAM arrays \cite{KingraXNOR22}. The SEM cross-section of fabricated RRAM device integrated on top of 130 nm CMOS technology is shown in the inset of Fig.~\ref{Fig1}(a). The device stack has TiN as BE (Bottom Electrode), non-stoichiometric SiO$_x$ as switching layer and TiN as TE (Top Electrode). The 1T-1R bitcell occupies 30F$^2$ on-chip area.  
Fig.~\ref{Fig1}(a,b) shows the electro-forming characteristics (where an initial conductive filament is formed) and cumulative distribution for pristine- and post-forming device resistance (in k$\Omega$). Fig.~\ref{Fig1}(c) shows SET and RESET switching characteristics highlighting D2D variability for the 8$\times$8 RRAM device array. LRS (Low Resistance State) and HRS (High Resistance State) device resistance distributions are shown in Fig.~\ref{Fig1}(d). It is observed that the LRS ranges from 3 k$\Omega$ to 20 k$\Omega$ and HRS ranges from 110 k$\Omega$ to 1 M$\Omega$. 
\begin{figure}[tb]
  \centering
  \includegraphics[width=0.75\linewidth]{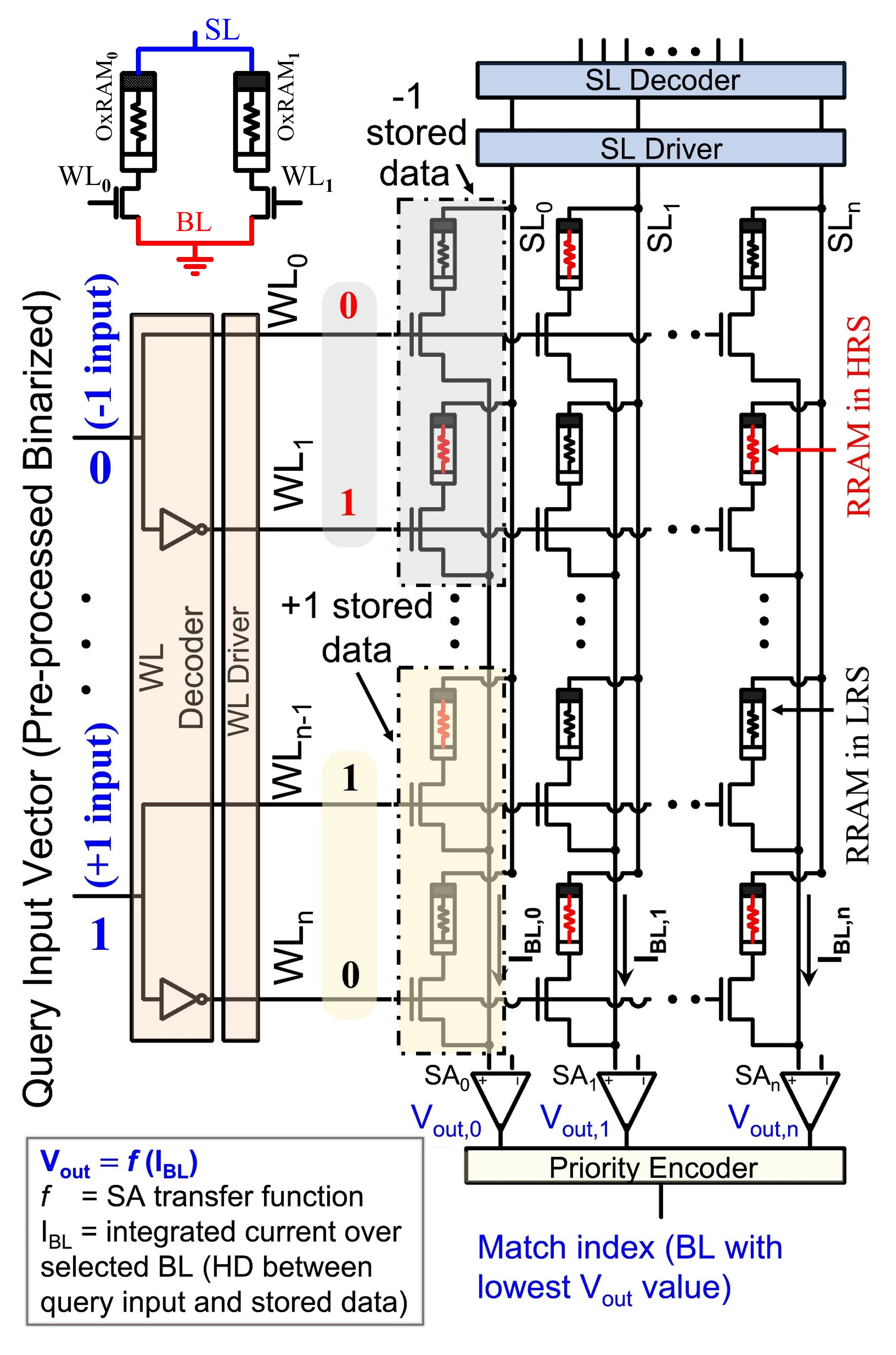}
    \vspace{-1em}
  \caption{Block diagram of the proposed binarized IMSS engine with periphery circuit blocks showing mapping of inputs for 2T-2R XOR bitcell. WL decoder maps QI vector in differential encoding. Current integrated along the bitlines ($I_{BL,n}$) is translated to voltage using SA to compute HD between applied QI vector and the corresponding SD vector. Inset at top left shows the circuit for 2T-2R XOR bitcell.}
  \label{Fig_xor}
\end{figure}
\begin{figure}[tb]
  \centering
  \includegraphics[width=0.85\linewidth]{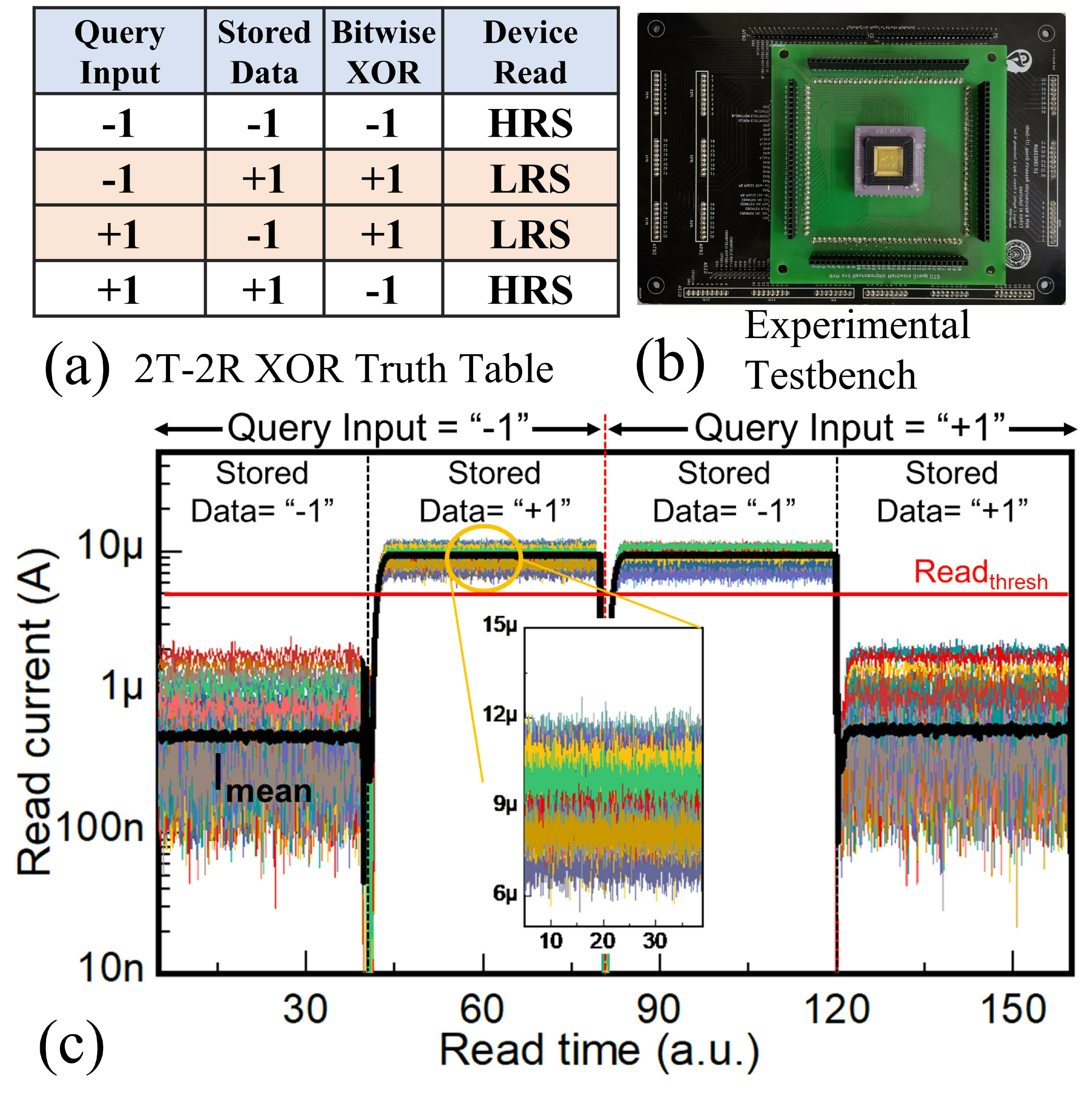}
    \vspace{-1em}
  \caption{(a) Truth table summarizing 2T-2R XOR bitcell operation, (b) Custom designed PCB for XOR IMC validation, (c) Experimental validation for all possible input QI combinations of 2T-2R XOR bitcell. }
  \label{Fig3}
\end{figure}

\section{RRAM IMSS: Experiments and Working}
\label{sec3}
{Schematic representation of 8$\times$8 1T-1R RRAM array used for experimental validation is illustrated in Fig.~\ref{Fig_xor}. To access a desired memory bitcell, the row address is selected by enabling corresponding WL and column address by selecting BL/SL. Data vectors are stored in the form of RRAM device conductance along columns (`-1' is encoded as top RRAM = LRS, bottom RRAM = HRS. While `+1' is encoded as top RRAM = HRS, bottom RRAM = LRS).
For programming RRAM to LRS, 1 $\mu$s long SET pulse is applied with V$_{WL}$=1.8 V, V$_{SL}$=1.4 V and V$_{BL}$=0 V. For HRS, 1 $\mu$s RESET pulse is applied with V$_{WL}$=4.5 V, V$_{SL}$=0 V and V$_{BL}$=1.2 V. To read bitcell resistance, a 50 $\mu$s READ pulse with V$_{WL}$=1.4 V, V$_{SL}$=0.2 V and V$_{BL}$=0 V is applied. For realizing a XOR gate in hardware, a 2T-2R bitcell is effectively realized by selecting two consecutive 1T-1R bitcells in the same column. QI is applied in binary format (`-1',`+1'). To eliminate the need for negative voltage (i.e., to represent `-1'), binary input is converted to a differential representation: `-1' $\rightarrow$ [0,1], and `+1' $\rightarrow$ [1,0] using a WL-decoder circuit. To perform XOR operation, SL is charged to 0.2 V and QI is applied as input to corresponding 2T-2R bitcell. Output of the circuit is obtained in the form of current flowing through corresponding BL. When the QI matches SD, RRAM in HRS is selected and negligible current flows. In case of a mismatch, RRAM in LRS is selected leading to higher output current. For a single column, output current of all XOR cells can be integrated following the principle of KCL (Kirchoff's Current Law) representing the HD between QI and SD as shown below in Eq. (\ref{eq1}). 
\begin{equation}
    HD(SD,QI)=\sum\limits_{i=0}^n QI[i] \oplus SD[i] \label{eq1}
\end{equation}}
Fig.~\ref{Fig3}(a) presents the truth table validating XOR gate functionality. In addition to XOR circuit functionality, the proposed scheme also improves robustness to programming errors as opposed to a single 1T-1R bitcell\cite{Bocquet_2018} due to differential storage. Fig.~\ref{Fig3}(b) shows the custom experimental setup and RRAM test chip used in the study. Programming signals are applied using high speed pulse measurement unit (Keithley 4225-PMU) from semiconductor parameter analyzer (Keithley 4200-SPA). The signals from PMU channels are multiplexed and applied to different signal lines (WL,SL,BL) using the custom switch board. Experimental validation of a single 2T-2R bitcell based XOR gate for all possible {input combinations across 32 2T-2R XOR bitcells} is presented in Fig.~\ref{Fig3}(c). Average $I_{read}$ (Read Current) is found to be less than 1 $\mu$A (i.e. RRAM device in HRS gets selected) when QI matches SD. When QI doesn't match SD i.e mismatch, average is $I_{read}$ $\geq$ 6 $\mu$A (i.e. RRAM device in LRS gets selected). A reliable sense-margin $\geq$ 5$\mu$A is realized between match and mismatch states. For a given QI vector, the current will be the lowest through BL storing match/nearest match data vector.} To perform IMSS, $V_{read}$ = 0.2 V is applied on SLs with SD vectors to be compared against applied QI vector. HD is computed by applying an \textit{n}-bit long binary QI vector and comparing it against \textit{n}-bit binary SD vectors; at each SL, there are \textit{n} 2T-2R XOR gates participating. Bit level comparison is carried out by each XOR gate and summed up I$_{BL}$ (Bit-line Current) from \textit{n} XOR gates is sensed on corresponding BL. BL/SL with maximum match bits will result in the lowest current. I$_{BL}$ is converted to V$_{out}$ (voltage sensed by SA (Sense Amplifier)).

\begin{figure*}[tb]
  \centering
  \includegraphics[width=0.95\linewidth]{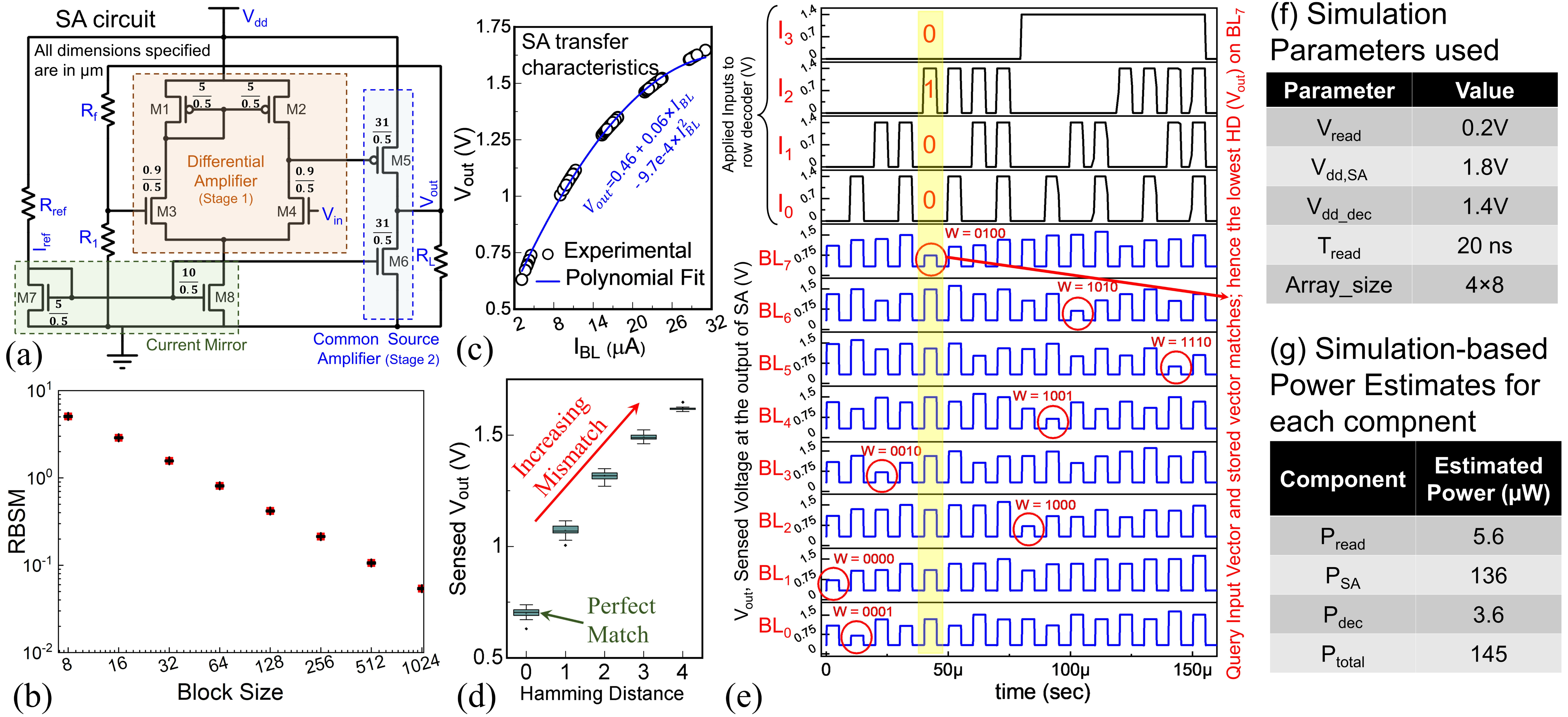}
  \vspace{-1em}
  \caption{(a) Schematic of the simulated SA circuit using two-stage amplification. (b) Impact of array sizing on sensing margin. RBSM is resistance-based sensing margin representing ratio in terms of R (in dB) for all-match and 1-mismatch case for the given block size. (c) SA transfer characteristic based on 130 nm Skywater PDK technology used for converting column-wise integrated current to voltage. (d) Simulations based SA outputs for different HD values using 4$\times$8 2T-2R array. (e) Timing waveforms for QI vector (applied at WLs) and V$_{out}$ for different BLs (storing binarized feature vectors) demonstrates successful match operation when QI vector matches with SD vector (extracted from SPICE simulations using 130 nm Skywater PDK). (f) Parameters used for SPICE simulations. (g) Estimated power values for each component.}
  \label{Fig_xor1}
\end{figure*}

\section{SPICE Simulations}
\label{sec4}
To validate the proposed IMSS architecture, SPICE simulations using the Opensource Skywater 130 nm PDK\cite{skywater} are performed with \textit{ngspice}. Simulations included RRAM arrays (modeled through resistance matrices) and periphery circuits shown in Fig.~\ref{Fig_xor} and Fig.~\ref{Fig_xor1}(a). Decoder-circuit/logic blocks and SA-circuits are designed with $V_{DD}$=1.4 V and 1.8 V respectively. Experimentally measured resistance values of LRS/HRS are used for the 1T-1R array during simulations. Transfer curve for the SA used in the simulation analysis is shown in Fig.~\ref{Fig_xor1}(c). Gain of SA is selected based on length of the column vector so that full-mismatch leads to V$_{out}$ $\approx$ V$_{DD}$ and full-match leads to V$_{out}$ $\approx$ 0 V. V$_{out}$ as a function of HD, between a 4-bit QI vector and a 4-bit SD vector is shown in Fig.~\ref{Fig_xor1}(d). Timing waveform showcasing all possible QI vector combinations and SA outputs for a 4$\times$8 2T-2R array are shown in Fig.~\ref{Fig_xor1}(e). V$_{out}$ is minimum when QI vector matches the SD vector. For instance, when applied QI is ``0100", V$_{out}$ for BL$_7$ is minimum because the SD along BL$_7$ is ``0100". After incorporating measured D2D variability (Fig.~\ref{Fig1}(b),(d)) in the simulations, the proposed IMSS scheme shows no overlap between output voltage levels for neighbouring HD values thus demonstrating a reliable operation. To estimate energy dissipation, current waveforms for all supply voltages ($I_{dd\_dec}$, $I_{read}$, $I_{dd\_SA}$) over a set of 32 input combinations were obtained through simulations. The simulations utilized a clock of 50 MHz. Based on average current dissipation for all operations, average power was estimated. Total power dissipation was estimated to be 145 $\mu$W with energy cost per XOR operation of 17 fJ. A complete breakdown of power estimation is shown in Fig.~\ref{Fig_xor1}(g) with methodology for estimation explained in Eq. (\ref{eq24})-(\ref{eq25}). In this study, we have exclusively focused on array size = 8$\times$8 in order to align with experimental measurements. However, array size has a significant contribution in determining overall performance for the IMC arrays\cite{KingraXNOR22}. For a constant workload, with increase in array size latency would decrease due to reduction in number of operation. However the cost of a single operation increases due to periphery overhead. Additionally increasing array size may lead to limitations in terms of sensing margin (see Fig.~\ref{Fig_xor1}(b)). 

\begin{align}
    P_{total} &= P_{read} + P_{SA} + P_{dec} \label{eq24}\\
    E_{XOR} &= \frac{P_{total} \times T_{read}}{Array\_size} \label{eq25}
\end{align}

\begin{algorithm}[t]
\begin{algorithmic}
\footnotesize{
\REQUIRE{Query Input vector QI, Stored Data SD} 
\ENSURE{Match Index m}
\STATE{\textbf{Pre-processing:}} 
\STATE{$q_1$ = PCA(QI) [0:20]} 
\STATE{$q_2$ = sign($q_1$) $\times$ $log_{10}$($q_1$)} 
\STATE{$q_3$ = $\frac{q_2 - \mu_1}{\sigma _1}$}
\STATE{$q_4$ = $\frac{q_3 - min_2}{max_2 - min_2}$}
\STATE{$q_5$ = $round(q_4 \times 255)$}
\FOR{i=0; i$<$8; i=i++}
    \STATE{$q_x$[i] = $q_5>(31 + 32 \times i)$}
\ENDFOR
\STATE{\textbf{Similarity Search:}} 
\FOR{k=0; k$<$len(SD); k=k++}
    \STATE{$d_x$[k] = $popcount(q_x \oplus SD[k])$}
\ENDFOR
\STATE{$m$ = $index(mode(topk(-d_x)))$}
}
\end{algorithmic}
\caption{Proposed IMSS method based on bitwise XOR.}
\label{algo1}
\end{algorithm}
\begin{figure*}[tb]
  \centering
  \includegraphics[width=0.9\linewidth]{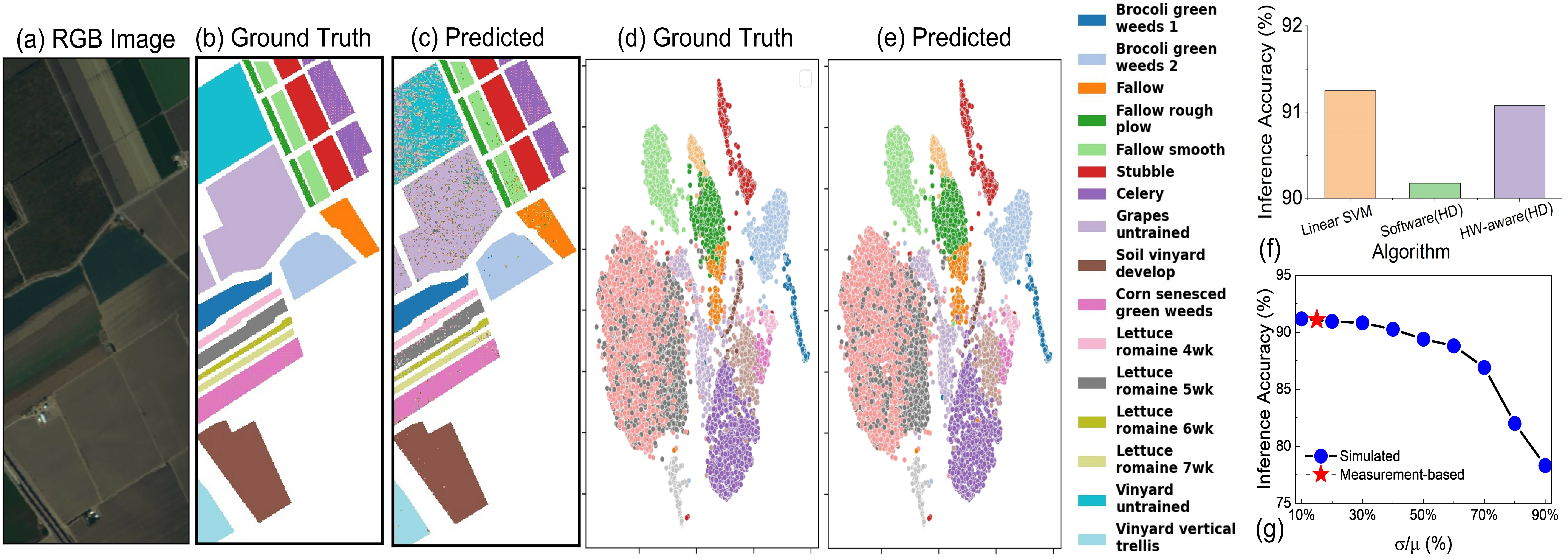}
  \vspace{-1em}
  \caption{HSI classification results using proposed IMSS engine. (a) RGB data, (b) Ground truth of classification, (c) Prediction results using proposed IMSS engine. Decision space representation based on t-SNE for: (d) Ground truth and (e) Predicted results. (f) Accuracy comparison for Linear SVM (Support Vector Machine), proposed algorithm and HW-aware simulations. (g) Impact of variability on performance of proposed scheme with variability expressed in terms of ratio to mean for device state (LRS/HRS). }
  \label{Fig4}
\end{figure*}

\begin{table*}[!htb]
  \centering
  \caption{Comparison of IMSS realized using structures with 2T-2R circuits.}
    \begin{tabular}{|c|c|c|c|c|c|c|c|c|c|}
    \hline
    \multirow{2}{*}{Ref.} & Device & Set  & Reset  & Read  & $E_{search}$ (pJ) & Tech.  &  \multirow{2}{*}{Array} & \multirow{2}{*}{Application} & \multirow{2}{*}{Dataset} \\
     & Stack &  Pulse &  Pulse & Pulse  & (128$\times$32) & Node &  &  &  \\

    \hline
    This & $SiO_x$ & 2V, 1$\mu$s & -1.4V, 2$\mu$s & 0.2V, 20ns   & 71.26 & 130 nm& \multirow{2}{*}{16 kB} & \multirow{2}{*}{Similarity Search}   & \multirow{2}{*}{Salinas} \\
    \cline{2-7}
     work & Si-doped $HfO_x$\cite{Grenouillet16k21} & 2.7V, 1$\mu$s & -2.7V, 1$\mu$s & 0.1V, 10ns & 28.67  & 28 nm &   &  &  \\
    \hline
    \cite{Li_2021} & $HfO_x$ & 3.3V, 1$\mu$s & -3.5V, 100$\mu$s & 0.2V, 10ns & 42.76  & 40 nm&  64 kB & One-shot learning & Omniglot \\
    \hline
    \cite{Yang_2019} & $MoS_2$ Tran. + $HfO_x$ & 2V, 0.2$\mu$s & -3V, 0.2$\mu$s & 50mV, 100ns & 0.82  & 90 nm&  Sim. & Similarity Search & NA \\
    \hline
    \cite{Ly_2019} & $HfO_x$ & 2V, 0.1$\mu$s & -2.5V, 0.1$\mu$s & 0.6V, 90ns & 44.46  & 130 nm&  4 kB & NA & NA \\
    \hline
    \cite{Ni_2019} & FeFET & -4V, 10$\mu$s & 4V, 10$\mu$s & 1V, 1ns & 4.62  & 45 nm&  Sim. & One-shot learning & Omniglot \\
    \hline
    \cite{Li_2014} & PCM & \multicolumn{2}{c|}{2.5V,10ns}  & 1.2V, 1.9ns & 32.13  & 90 nm&  1 Mb & Similarity Search & NA \\
    \hline
    \cite{Matsunaga_2014} & MRAM & 50$\mu$A & -150$\mu$A & 1.5V & NA  & 140 nm &  9kb & Similarity Search & NA \\
    \hline

    \end{tabular}%
  \label{tab1}%
\end{table*}%


\section{HSI Classification Task}
\label{sec5}
HSI classification on the `Salinas' dataset is used as an example usecase for the proposed IMSS architecture where every pixel in the image is classified to identify the type and age of the vegetation present on ground. Dataset includes HSI of size 512$\times$217 with 224 bands acquired using AVIRIS (Airborne Visible/Infrared Imaging Spectrometer) \cite{vane1993airborne} sensor flying over Salinas Valley, California. RGB representation of the image is shown in Fig.~\ref{Fig4}(a). Ground truth comprises of 16 types of vegetation classes (see Fig.~\ref{Fig4}(b)). Design of HSI application-specific pre-processing pipeline becomes essential in order to obtain a simplified representation of the data without losing information. In case of HSI since both data precision and spectral resolution is high, compression becomes important. Custom pre-processing steps used for performing IMSS using Salinas dataset are summarized in Algorithm~\ref{algo1}. First step involves performing PCA (Principal Component Analysis) \cite{wold1987principal} to extract relevant feature data, followed by log-scaling to compress data representation especially in case of large integers. To preserve the dynamic range, sign-multiplication is performed followed by normalization using two methods: mean-sigma followed by min-max. At the final step, 8-bit unsigned integer (\textit{uint8}) representation is created. In the proposed HSI classification pipeline, use of only XOR operations is ensured to exploit the 2T-2R IMC array. Due to position specific weight assigned to each bit in case of fixed-point numbers, directly computing HD may lead to inaccurate match operation. To overcome this issue, \textit{uint8} values are converted to an 8-bit thermometric encoding \cite{buckman2018thermometer} with a resolution of 32. This facilitates assigning equal numerical significance to each bit thus making HD a feasible metric for performing IMSS operation. Using the aforementioned pre-processing pipeline, input HSI data is compressed to 20 channels where each channel is translated to 8 channels of binary thermometric encoding. Each pixel in the original image is stored as a 160 bit vector (20$\times$8) i.e. compression by a factor of $\approx$44$\times$. For training, 70\% of the dataset (image pixels) is used. Entire training data, based on the proposed differential encoding scheme, would require an RRAM IMSS chip (Fig.~\ref{Fig_xor}) of size $\approx$24MB. Since this requires a modest chip size for high-density RRAM, software simulations were performed using the PyTorch framework to validate the full Salinas classification application based on proposed RRAM IMSS. Pre-processed feature vectors from the test set are applied as QI to compute HD through XOR operations between SD vector and QI vector. Index of the SD vector with least distance is computed and class label from corresponding index is assigned to the test vector. Inference results using proposed IMSS scheme are shown in Fig.~\ref{Fig4}(c). To visualize learning performance in form of 2D-decision spaces, t-SNE (t-Distributed Stochastic Neighbour Embedding) \cite{vandermaaten08a} plots are generated for both ground-truth and predicted results as shown in Fig.~\ref{Fig4}(d) and Fig.~\ref{Fig4}(e) respectively. A comparison of the proposed HW-based approach against standard classification algorithms and software-based realization is shown in Fig.~\ref{Fig4}(f). For HW-aware simulations we incorporate the experimentally measured D2D variability of individual resistance states as shown in Fig.~\ref{Fig1}(d). To further demonstrate the resilience benefits of the 2T-2R structure, inference accuracies were estimated by sweeping the device variability for both HRS and LRS states simultaneously (see Fig.~\ref{Fig4}(g)). Table~\ref{tab1} presents comparison with other RRAM-based IMSS studies in literature. The proposed methodology enables simultaneous processing of 8 rows to determine HD as opposed to prior work where only 4 rows can be sensed in a single cycle of operation\cite{Li_2021}. This however comes at the cost of increased energy dissipation from the SA circuit. When comparing device read energy costs for processing IMSS across 32 vectors of 128-bit, it can be observed that current implementation consumes $\approx$2$\times$ more energy compared to  state-of-the-art array-based implementations. For the case of FeFET and MoS$_2$ transistor-based realizations, the array structures have been realized in simulation and hence haven't been considered for comparison. However, projected estimations using parameters from a 28 nm fabricated 1T-1R arrays\cite{Grenouillet16k21} demonstrates energy savings of $\approx$2.5$\times$ compared to present implementation and $\approx$1.5$\times$ compared to SOTA IMC bitcells. 



\section{Conclusion}
\label{sec6}
Successful realization of XOR operations based on 2T-2R circuits using fabricated $SiO_x$ RRAM was experimentally validated. Analog computation of HD based on column-wise current integration to perform binarized IMSS was realized through extensive SPICE simulations including peripheral circuits using the Skywater 130 nm PDK. Simulations validated energy of 17 fJ per XOR operation with a full-system power dissipation of 145 $\mu$W for 8$\times$8 RRAM array. Using projected estimations at advanced nodes (28 nm) energy savings of $\approx$1.5$\times$ compared to the state-of-the-art can be observed for a fixed workload. Proposed IMSS scheme was used for HSI pixel classification demonstrating 91\% accuracy.

\bibliographystyle{IEEEtran_mod}
\bibliography{ref}

\begin{thebibliography}{10}
\providecommand{\url}[1]{#1}
\csname url@samestyle\endcsname
\providecommand{\newblock}{\relax}
\providecommand{\bibinfo}[2]{#2}
\providecommand{\BIBentrySTDinterwordspacing}{\spaceskip=0pt\relax}
\providecommand{\BIBentryALTinterwordstretchfactor}{4}
\providecommand{\BIBentryALTinterwordspacing}{\spaceskip=\fontdimen2\font plus
\BIBentryALTinterwordstretchfactor\fontdimen3\font minus
  \fontdimen4\font\relax}
\providecommand{\BIBforeignlanguage}[2]{{%
\expandafter\ifx\csname l@#1\endcsname\relax
\typeout{** WARNING: IEEEtran.bst: No hyphenation pattern has been}%
\typeout{** loaded for the language `#1'. Using the pattern for}%
\typeout{** the default language instead.}%
\else
\language=\csname l@#1\endcsname
\fi
#2}}
\providecommand{\BIBdecl}{\relax}
\BIBdecl

\bibitem{PanFSP21}
Y.~Pan \emph{et~al.}, ``A rram-based associative memory cell,'' in
  \emph{{ISCAS} 2021, Daegu, South Korea, May 22-28, 2021}.\hskip 1em plus
  0.5em minus 0.4em\relax {IEEE}, 2021, pp. 1--5.

\bibitem{ImaniMWPR19}
M.~Imani \emph{et~al.}, ``A binary learning framework for hyperdimensional
  computing,'' in \emph{{DATE}}.\hskip 1em plus 0.5em minus 0.4em\relax {IEEE},
  2019, pp. 126--131.

\bibitem{Wu_2018}
T.~F. Wu \emph{et~al.}, ``Hyperdimensional computing exploiting carbon nanotube
  fets, resistive ram, and their monolithic 3d integration,'' \emph{{IEEE}
  JSSC}, vol.~53, no.~11, pp. 3183--3196, 2018.

\bibitem{Imani_2020}
M.~Imani \emph{et~al.}, ``Dual: Acceleration of clustering algorithms using
  digital-based processing in-memory,'' in \emph{{MICRO}}, 2020, pp. 356--371.

\bibitem{Li_2021}
H.~Li \emph{et~al.}, ``Sapiens: A 64-kb rram-based non-volatile associative
  memory for one-shot learning and inference at the edge,'' \emph{IEEE {TED}},
  pp. 1--7, 2021.

\bibitem{MohanS09}
N.~Mohan \emph{et~al.}, ``Low-leakage storage cells for ternary content
  addressable memories,'' \emph{{IEEE} {TVLSI}}, vol.~17, no.~5, pp. 604--612,
  2009.

\bibitem{PagiamtzisS06}
K.~Pagiamtzis \emph{et~al.}, ``Content-addressable memory {(CAM)} circuits and
  architectures: a tutorial and survey,'' \emph{{IEEE} {JSSC}}, vol.~41, no.~3,
  pp. 712--727, 2006.

\bibitem{Noda_2005}
H.~Noda \emph{et~al.}, ``A cost-efficient high-performance dynamic {TCAM} with
  pipelined hierarchical searching and shift redundancy architecture,''
  \emph{{IEEE} {JSSC}}, vol.~40, no.~1, pp. 245--253, jan 2005.

\bibitem{Lines}
V.~Lines \emph{et~al.}, ``66 {MHz} 2.3 m ternary dynamic content addressable
  memory,'' in \emph{Records of the {IEEE} International Workshop on Memory
  Technology, Design and Testing}.\hskip 1em plus 0.5em minus 0.4em\relax
  {IEEE} Comput. Soc, 2000.

\bibitem{7159147}
R.~Karam \emph{et~al.}, ``Emerging trends in design and applications of
  memory-based computing and content-addressable memories,'' \emph{Proceedings
  of the IEEE}, vol. 103, no.~8, pp. 1311--1330, 2015.

\bibitem{Sun_2018}
X.~Sun \emph{et~al.}, ``Xnor-rram: A scalable and parallel resistive synaptic
  architecture for binary neural networks,'' in \emph{DATE}, 2018, pp.
  1423--1428.

\bibitem{Shreya_2020}
S.~Shreya \emph{et~al.}, ``Energy-efficient all-spin bnn using
  voltage-controlled spin-orbit torque device for digit recognition,''
  \emph{IEEE TED}, vol.~68, no.~1, pp. 385--392, 2021.

\bibitem{Ielmini_2018}
D.~Ielmini \emph{et~al.}, ``In-memory computing with resistive switching
  devices,'' \emph{Nature Electronics}, vol.~1, no.~6, pp. 333--343, jun 2018.

\bibitem{Matsunaga_2014}
S.~Matsunaga \emph{et~al.}, ``Implementation of a perpendicular mtj-based
  read-disturb-tolerant 2t-2r nonvolatile tcam based on a reversed current
  reading scheme,'' in \emph{ASPDAC}, 2012, pp. 475--476.

\bibitem{Li_2014}
J.~Li \emph{et~al.}, ``1 mb 0.41 µm² 2t-2r cell nonvolatile tcam with two-bit
  encoding and clocked self-referenced sensing,'' \emph{IEEE JSSC}, vol.~49,
  no.~4, pp. 896--907, 2014.

\bibitem{Ni_2019}
K.~Ni \emph{et~al.}, ``Ferroelectric ternary content-addressable memory for
  one-shot learning,'' \emph{Nature Electronics}, vol.~2, no.~11, pp. 521--529,
  nov 2019.

\bibitem{Yang_2019}
R.~Yang \emph{et~al.}, ``Ternary content-addressable memory with {MoS}2
  transistors for massively parallel data search,'' \emph{Nature Electronics},
  vol.~2, no.~3, pp. 108--114, mar 2019.

\bibitem{HalawaniLMAA18}
Y.~Halawani \emph{et~al.}, ``Stateful memristor-based search architecture,''
  \emph{{IEEE} {TVLSI}}, vol.~26, no.~12, pp. 2773--2780, 2018.

\bibitem{Bocquet_2018}
M.~Bocquet \emph{et~al.}, ``In-memory and error-immune differential rram
  implementation of binarized deep neural networks,'' in \emph{IEDM}, 2018, pp.
  20.6.1--20.6.4.

\bibitem{KingraXNOR22}
S.~K. Kingra \emph{et~al.}, ``Dual-configuration in-memory computing bitcells
  using siox rram for binary neural networks,'' \emph{Applied Physics Letters},
  vol. 120, no.~3, p. 034102, 2022.

\bibitem{skywater}
``Skywater open source pdk,'' \url{https://github.com/google/skywater-pdk}.

\bibitem{Grenouillet16k21}
L.~Grenouillet \emph{et~al.}, ``16kbit 1t1r oxram arrays embedded in 28nm fdsoi
  technology demonstrating low ber, high endurance, and compatibility with core
  logic transistors,'' in \emph{IMW}, 2021, pp. 1--4.

\bibitem{Ly_2019}
D.~R.~B. Ly \emph{et~al.}, ``Novel 1t2r1t rram-based ternary content
  addressable memory for large scale pattern recognition,'' in \emph{IEDM},
  2019, pp. 35.5.1--35.5.4.

\bibitem{vane1993airborne}
G.~Vane \emph{et~al.}, ``The airborne visible/infrared imaging spectrometer
  (aviris),'' \emph{Remote sensing of environment}, vol.~44, no. 2-3, pp.
  127--143, 1993.

\bibitem{wold1987principal}
S.~Wold \emph{et~al.}, ``Principal component analysis,'' \emph{Chemometrics and
  intelligent laboratory systems}, vol.~2, no. 1-3, pp. 37--52, 1987.

\bibitem{buckman2018thermometer}
J.~Buckman \emph{et~al.}, ``Thermometer encoding: One hot way to resist
  adversarial examples,'' in \emph{ICLR}, 2018, pp. 1--22.

\bibitem{vandermaaten08a}
L.~van~der Maaten \emph{et~al.}, ``Visualizing data using t-sne,'' \emph{JMLR},
  vol.~9, no.~86, pp. 2579--2605, 2008.

\end{thebibliography}
\end{document}